# PROBABILISTIC SVM/GMM CLASSIFIER FOR SPEAKER-INDEPENDENT VOWEL RECOGNITION IN CONTINUES SPEECH

*Mohammad Nazari*[a], *Abolghasem Sayadiyan*[a], *SeyedMajid Valiollahzadeh*[b]

[a]Department of Electrical Engineering, Amirkabir University of Technology, Tehran, Iran
[b]Department of Electrical Engineering, Sharif University of Technology, Tehran, Iran

**ABSTRACT**

In this paper, we discuss the issues in automatic recognition of vowels in Persian language. The present work focuses on new statistical method of recognition of vowels as a basic unit of syllables. First we describe a vowel detection system then briefly discuss how the detected vowels can feed to recognition unit. According to pattern recognition, Support Vector Machines (SVM) as a discriminative classifier and Gaussian mixture model (GMM) as a generative model classifier are two most popular techniques. Current state-of-the-art systems try to combine them together for achieving more power of classification and improving the performance of the recognition systems. The main idea of the study is to combine probabilistic SVM and traditional GMM pattern classification with some characteristic of speech like band-pass energy to achieve better classification rate. This idea has been analytically formulated and tested on a FarsDat based vowel recognition system. The results show inconceivable increases in recognition accuracy. The tests have been carried out by various proposed vowel recognition algorithms and the results have been compared.

## 1. INTRODUCTION

Some recent researches in speech processing area like Automatic Speech Recognition (ASR) and speaker recognition and verification focus on Vowel Recognition (VR) because of generally spectrally well defined character of vowels. In fact they improve our ability to recognize speech significantly, both by human beings and ASR systems. Therefore the vowel recognition generally plays an important role [1, 2 and 3].

Previous vowel recognition methods like segmental trajectory modeling [3] and HMM using cepstra with their derivatives have some leakage. For example HMM can not model the trajectories of speech signals effectively especially for vowels. In segmental trajectory modeling, the main problem is computational complexity of estimation of transformation matrix to reduce the high correlation within the residual error covariance using Minimum Classification Error (MCE).

In this paper, like traditional segmental modeling methods [5, 6], we proposed a weighted least square estimation to estimate the trajectory feature but for reducing the computational complexity we weaken the updating of transformation matrix then we used the state-of-the-art maximum margin classifier, Probabilistic Vector Machines (PSVM) [9, 10], as a powerfully discriminative function to compensate lack of accuracy. SVM is an effective and accurate discriminative model and it has excellent property of making full use of discriminative information of different classes in the representation pattern variations.

Generative model such as Gaussian Mixture Model (GMM) can construct high performance class models for pattern recognition tasks using statistical information. Earlier works try to combine generative models, particularly GMMs and HMMs, with discriminative framework like SVM [7]. In these systems classifiers are trained to discriminate between individual frames of data then the likelihood scores of each frame are combined using an averaging step [8] to give an overall utterance score from which the authenticity of the speaker may be determined. In this paper we introduce a solution to combine weighted GMM for selecting the SVM training data set to prevail over an important weakness of SVM in large scale databases. Therefore we use GMM score (likelihood) for classifying the easy-to-find members of classes and keep other hard-to-find members, we can reduce number of support vectors.

The rest of the paper is organized as follows. In Section two, we start the analysis by briefly reviewing the SVM and GMM. Following that in section three, we introduce our method as a powerful classifier. In Section four, a set of experiments are presented to demonstrate the effectiveness of our classifier. The proposed method is compared, in terms of the classification error rate performance, to other methods like pure SVM for Speaker-Independent Vowel Recognition on the FarsDat speech database. Conclusions are summarized in Section five.

## 2. SVM CLASSIFIER WITH GMM TRAINING SET SELECTION

### 2.1. Support Vector Machines (SVM)

The principle of Support Vector Machine (SVM) relies on a linear separation in a high dimension feature space where the data have been previously mapped considering the eventual non-linearities. Assuming that the training set $X = (x_i)_{i=1}^l \subset R^R$ is labeled with two class targets $Y = (y_i)_{i=1}^l$ with $l$ the number of training vectors, R the real line and R number of modalities:

$$y_i \in \{-1, +1\} \quad \Phi : R^R \to F \quad (1)$$

Y, Maps the data into a feature space F. it has been proved that maximizing the minimum distance in space F between $\Phi(X)$ and the separating hyper plane $H(w,b)$ is a good means of reducing the generalization risk [10].

$$H(w,b) = \{f \in F \mid <w, f>_F + b = 0\} \quad (2)$$

Where, $<>$ is inner product Also, it has been proved that the optimal hyper plane can be obtained solving the convex quadratic programming (QP) problem [10]:

$$\text{Minimize} \quad \frac{1}{2}\|w\|^2 + c\sum_{i=1}^l \xi$$
$$\text{with} \quad y_i(<w, \Phi(X)> + b) \geq 1 - \xi \quad i = 1,...,l \quad (3)$$

Where constant C and slack variables x are introduced to take into account the eventual non-separability of $\Phi(X)$ into F. Practically, this criterion is softened to the minimization of a cost factor involving both the complexity of the classifier, the degree to which marginal points are misclassified, and the tradeoff between these factors through a margin of error parameter (usually designated C) which is tuned through cross-validation procedures. There are several common kernel functions that are used such as the linear, polynomial kernel, sigmoidal kernel and the most popular one, Gaussian (or "radial basis function") kernel, defined as:

$$K(x, y) = \exp(\frac{-|x-y|^2}{(2\sigma^2)}) \quad (4)$$

Where $\sigma$ is a scale parameter and x, y are feature-vectors in the input space. The Gaussian kernel has two hyper parameters of C and $\sigma$ to control the overall performance. In this paper we used radial basis function (RBF).

### 2.2. Probabilistic SVM

Given training examples $\hat{x}_i \in \Re^n, i = 1,...,m$, labeled by $\hat{y}_i \in \{+1, -1\}$, the binary Support Vector Machine (SVM) computes a decision function $f(x)$ such that $sign(f(x))$ can be used to predict the label of any test example $x$. Instead of predicting the label, many applications require a posterior class probability $P(y = 1 \mid x)$. Platt [9] proposes to approximate the posterior by a sigmoid function:

$$P(y = 1 \mid x) \approx P_{A,B}(x) \equiv \frac{1}{1 + \exp(Af(x) + B)} \quad (5)$$

The best parameters $(A, B)$ are then estimated by solving the following regularized maximum likelihood problem with a set of labeled examples $\{(x_i, y_i)\}_{i=1}^l$ (with $N_+$ of the $y_i$'s positive and $N_-$ for negative ones):

$$\min_{z=(A,B)} F(z) = -\sum_{i=1}^l (t_i \log(p_i) + (1 - t_i) \log(1 - p_i)), \quad (6)$$

$$\text{for } p_i = P_{A,B}(x_i), \text{ and } t_i = \begin{cases} \frac{N_+ + 1}{N_+ + 2} & \text{if } y_i = +1 \\ \frac{1}{N_- + 2} & \text{if } y_i = -1 \end{cases}, \quad i = 1,...,l$$

### 2.3. Gaussian Mixture Models

Gaussian Mixture Models (GMM) provides a good approximation of the originally observed feature probability density functions by a mixture of weighted Gaussians. The mixture coefficients were computed using an Expectation Maximization (EM) algorithm. Each emotion is modeled in a separate GMM and decision is made on the basis of maximum likelihood model. We used diagonal covariance GMMs as baseline classifier. The outputs of GMM are:

$$P_{GMM}(x \mid C_i) = \sum_{m=1}^M c_{im} N(x, \mu_{im}, \Sigma_{im}) \quad (7)$$

Where:

$$N(x, \mu_{im}, \Sigma_{im}) = \frac{1}{(2\pi)^{d/2}|\Sigma|^{1/2}} \times \exp\left[-\frac{1}{2}(x-\mu)^T \Sigma^{-1}(x-\mu)\right] \quad (8)$$

Here, $c_{im}$, $\mu_{im}$ and $\Sigma_{im}$ are the weight, mean and variance, respectively, of the m-th mixture for class i. The GMM reflects the intra-class information.

## 3. VOWEL RECOGNITION SYSTEM OVERVIEW

To make a model practical, it is necessary to develop training and recognition algorithms precisely. Our system based on two important steps, first step is vowel detection and second for vowel classification. In following we briefly describe our steps.

### 3.1. Vowel detection and recognition

The purpose of this step is creation of system for detection of vowels then finds the best boundaries of vowels. In fact this step is a pre-processing for classification step. Outputs of vowel detection block include two boundaries (start and end point of vowel) and average likelihood score of each vowel's segment. The basis of the suggested model is a linear fusion of estimated score of GMM's with probabilistic SVM and traditional band-pass energy for achieving better performance and accuracy. In rest of this section we describe proposed vowel recognition system.

## 3.2. Soft GMM Fitting

The main idea of soft segment modeling on a phoneme recognition system is proposed in [5] and improved in [6]. In this segmentation method, considers neighbor segment's vectors in estimating each segment's probability distribution function (PDF) with suitable weight using a GMM. The importance of soft segmentation approach may come into view in the boundary estimation and the recognition phase. In the training phase, the adjacent segments are playing role in GMM parameter estimation.

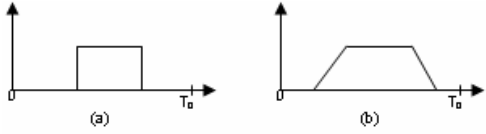

Figure 1. Soft segment modeling versus hard segment modeling. (a)Hard (b) Soft segment modeling.

We proposed to compute this score (normalized between 0 and 1), called confidence measure (CM), to indicate reliability of any recognition decision made by vowel detection system. CM can be computed for every recognized vowel to indicate how likely it is correctly recognized and how much we can trust the results for the utterance.

## 3.3. Probabilistic SVM Training with GMM's Output

The overall training and recognition block diagram of the developed system is depicted in figure 2. In this proposed method, we introduce a method that how we can use GMM for selecting the SVM training data set. An important weakness of SVM in large scale databases is time consuming in real time recognition because of its large numbers of support vectors.

In this case, if we use GMM confidence measure for choosing the training dataset, we achieve the best support vectors. We discuss vowel ($\omega_1$) and non-vowel ($\omega_2$) training system, The GMM score is the difference between the log likelihoods of the two models,

$$l(X) = \log P(X \mid \omega_1) - \log P(X \mid \omega_2) \qquad (9)$$

The decision boundary is:

$$l(X) \underset{\omega_2}{\overset{\omega_1}{\underset{>}{<}}} \log(P_1) - \log(P_2) \qquad (10)$$

Where $P_i$ is the A priori probability of $\omega_i$. If we add $\varepsilon$ margin for GMM score,

$$\begin{cases} X \in \omega_1, & if \quad l(X) < \log\left(\frac{P_1}{P_2}\right) - \varepsilon \\ X \in \omega_2, & if \quad l(X) > \log\left(\frac{P_1}{P_2}\right) + \varepsilon \\ X \text{ pass to SVM classifier} & if \quad otherwise \end{cases} \qquad (11)$$

That $\varepsilon$ is calculated experimentally. It is clear that the value of $\varepsilon$ is very important for generalization characteristic of classifier.

## 3.4. Classification with linear combined models' outputs

In this section, we proposed a linear model for vowel recognition based on combining the outputs of soft GMM models (vowel and non-vowel classes), the probabilistic output of PSVM and band-pass energy. The overall training block diagram of the developed system is depicted in figure 1. We suggested for calculation of vowel boundaries first we must estimate $P(Vowel \mid X)$:

$$P(Vowel \mid X) \cong (0.3)G(X) \\ + (0.5)P_{GMM}(X \mid Vowel) + (0.2)P_{PSVM}(X) \qquad (12)$$

Where $P(Vowel \mid X)$ is probability of input vector $X$ is member of vowel class, $P_{GMM}(X \mid Vowel)$ is output of soft GMM fitted to vowel class, $P_{PSVM}(X)$ is output of PSVM and $G(X)$ is band-pass energy of frame. Like vowel class we can calculate $P(NonVowel \mid X)$ for non vowel class:

$$P(NonVowel \mid X) \cong (0.3)(1 - G(X)) \\ + (0.5)P_{GMM}(X \mid NonVowel) + (0.2)P_{PSVM}(X) \qquad (13)$$

Where $P_{GMM}(X \mid NonVowel)$ is the output of soft GMM fitted to non-vowel class. The underlying goal of classifier combination theory is to identify the conditions under which the combination of an ensemble of classifiers yields improved performance compared to the individual classifiers. We can find the vowels boundaries with contact points of $P(Vowel \mid X)$ and $P_{GMM}(X \mid NonVowel)$ curves.

## 4. EXPERIMENTAL RESULTS

The proposed method has been verified on a subset of clean speech data consisting of 30 male and 25 female utterances with no background noise were extracted from FARSI-DAT (most popular Persian speech database) for the-evaluation experiments. The training material consisted of 30 complete sentences for each speaker. We made vowels database by labeling vowels manually in each utterance. Our database focuses on eight important vowels in Persian language (as /a/, /@/, /o/, /e/, /i/, /u/, /au/, /ei/). We used 80 percent of our data for training and 20 percent for evaluation phase. Specifications of the speech analysis at the acoustic pre-processor are summarized as follows (in Table 1):

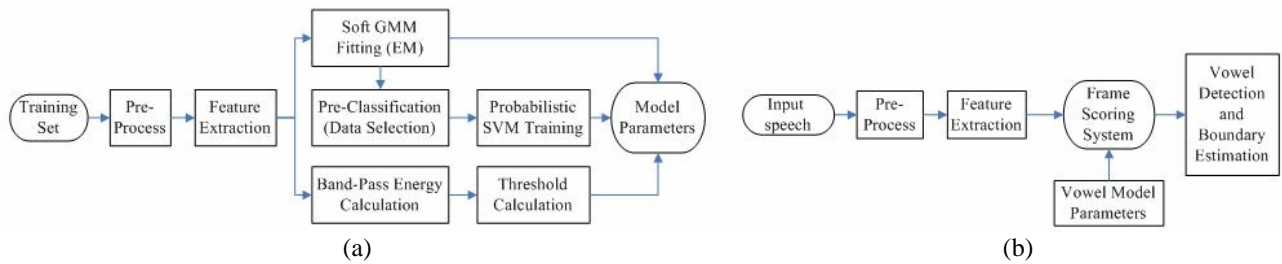
Figure 1 – Block Diagram of (a) Training phase (b) Vowel Recognition system.

In training phase, we searched for the best number of mixtures in soft GMM model experimentally. For vowel class, GMM have been trained with 80 mixtures and for non-vowel class, with 170 mixtures. Although this may increase the computational cost, it would be ignorable in comparison with Viterbi search computational cost.

In recognition phase, the results compared by equivalent system using HMM. In overall, our method improved about 1% in averaged recognition rate. Accuracy matrix for proposed speaker independent vowel recognition is illustrated in Table 2.

| Sampling Frequency | 8 kHz |
|---|---|
| Pre-Emphasis | $1 - 0.98\, z^{-1}$ |
| Hamming window width | 25 ms(200 Point) |
| Frame period | 12.5 ms(100 Point) |
| LPC analysis order | 16-th |
| Feature parameters | MFCC, Delta MFCC, Delta Log-Energy |

Table 1. Speech analysis conditions

|  |  | Uttered Vowel | | | | | | |
|---|---|---|---|---|---|---|---|---|
|  |  | /a/ | /@/ | /o/ | /e/ | /i/ | /u/ | /au/ |
| Recognized Vowel | /a/ | 94.9 | 2.1 | 0.3 | 0.5 | 0.1 | 0.1 | 0.1 |
| | /@/ | 1.4 | 95.2 | 0.1 | 0.8 | 0.3 | 0.4 | 0.2 |
| | /O/ | 0.8 | 0.4 | 96.1 | 3.7 | 0.4 | 0.2 | 0.4 |
| | /e/ | 0.4 | 0.5 | 2.1 | 93.2 | 0.3 | 0.1 | 0.1 |
| | /i/ | 0.5 | 1.1 | 0.3 | 0.4 | 95.1 | 1.1 | 1.4 |
| | /u/ | 0.9 | 0.4 | 0.7 | 0.6 | 2.9 | 96.1 | 0.6 |
| | /au/ | 1.1 | 0.3 | 0.4 | 0.8 | 0.9 | 2.1 | 97.2 |

Table 2. Accuracy matrix for Vowel Recognition system

## 5. DISCUSSIONS AND CONCLUSIONS

A simple and efficient statistical Vowel Recognition method has been introduced in this paper. This method improved accuracy of vowel recognition with combining GMM, SVM and Band-pass Energy. The main feature of this model is the toleration of gradual inter-segmental conversion. The model is very promising in both recognition rate and computational complexity aspects. The proposed method has the ability to reduce support vectors significantly. This reduction leads us to improve the speed of SVM classifier also using the GMM help us achieving more accuracy. The main advantage of this model is a drastic reduction of recognition time. The remained open problems are the soft window shape, fast methods for both GMM recognition and training, and the coefficients of each combined classifiers (e.g. SVM, GMM and Band-Pass Energy) on this duration modeling approach, which their studies are all in progress now.